\documentstyle[preprint,aps]{revtex}

\begin{document}
\input{epsf}
\draft
\title{COMPRESSIBILITY, CAPACITANCE AND GROUND STATE ENERGY FLUCTUATIONS\\
IN A WEAKLY INTERACTING QUANTUM DOT}

\author
{R. Berkovits}

\address{
The Jack and Pearl Resnick Institute of Advanced Technology,\\
Department of Physics, Bar-Ilan University,
Ramat-Gan 52900, Israel}

\author{B. L. Altshuler}

\address{
NEC Research Institute, 4 Independence Way, Princeton, NJ 08540}

\date{\today}
\maketitle

\begin{abstract}
We study the effect of electron-electron (e-e)
interactions on 
compressibility, capacitance and inverse compressibility of electrons in
a quantum dot or a small metallic grain. The calculation is performed
in the random-phase approximation. As expected,
the ensemble-averaged compressibility and capacitance
decreases as a function of the interaction
strength, while the mean inverse compressibility increases.
Fluctuations of the compressibility are found to be strongly suppressed
by the e-e interactions. Fluctuations of the capacitance and inverse 
compressibility also turn
out to be much smaller than their averages. 
The analytical calculations are compared with the results of a
numerical calculation for the inverse compressibility of a disordered
tight-binding model. Excellent agreement for weak interaction values 
is found. Implications for the
interpretation of current experimental data are discussed.
\end{abstract}

\section{Introduction}

The interplay between disorder and interactions is
one of the most exciting topics in mesoscopic physics.
Both factors play an important role in determining
the change of the number of electrons in a mesoscopic system as 
the chemical potential of the system is changed.
The investigation of this problem is
especially timely since ground-state
properties of disordered interacting systems
have recently become experimentally accessible. For example, by measuring 
the spacings between the gate voltages for which a quantum dot connected
to external leads exhibits conductance peaks, one can infer the
inverse compressibility of the electron gas in the dot
\cite{mk,ash,chang,marc}. 
Devices of this type have recently been used to measure the many-particle
ground-state energy as a function of the number of electrons in a dot
and large fluctuations (compared to the single electron level spacing) 
in this property were observed\cite{chang,marc,sba}. Large fluctuations
have also been observed in earlier experiments in which
the inverse compressibility 
of an insulating indium-oxide wire was measured \cite{cow}.

In the absence of interactions the derivative of the chemical potential
$\mu$ with respect to the number of electrons $N$
(i.e., the inverse compressibility, equal also to the second derivative 
with respect to $N$ of the
ground state energy) of a system equals
to the single-electron level spacing. It is well known that the
single-electron level statistics in disordered systems
are connected to the statistical
properties of random matrices \cite{ge,as,metha}. The level spacings exhibit
fluctuations and therefore one expects that the inverse compressibility 
of different samples, as well as the inverse compressibility
for the same sample at different chemical potentials,
will also exhibit fluctuations of order of the single-electron mean level
spacing $\Delta$. In the non-interacting
regime there are many theoretical approaches which can be used
to calculate these fluctuations, among them
are the random matrix theory (RMT)\cite{metha}, 
perturbative diagrammatic calculations
\cite{as} and the supersymmetry method \cite{efetov,vwz}.

On the other hand,
once electron-electron (e-e) interactions are included the situation becomes
more complicated. In contrast to the high 
energy excitations of an interacting system for which a RMT description
works very well (for example the high excitations of a nuclei\cite{metha}
and of disordered interacting systems\cite{ber} ),
The {\it ground-state} energy of an interacting system as a function of the
number of particles can not be understood by means of RMT. 

The average compressibility $\langle \partial N/\partial \mu \rangle$,
as well as the average capacitance $\langle C \rangle$,
are expected to decrease as result of interactions since
it costs more energy to add particles into the system.
This is somewhat similar to the situation for the average polarizability
suppression by the e-e interactions \cite{srw,rss}. 

It is not a priori clear 
how the interactions will influence the fluctuations.
Previously we have calculated
the influence of e-e interactions on the sample-to-sample fluctuations
of the polarizability \cite{ba}. 
By analogy one might expect that the relative magnitude 
of the compressibility and capacitance fluctuations should be suppressed. 

The average inverse compressibility $\langle \partial \mu/\partial N \rangle$
is expected to increase as a function of interaction strength
for repulsive interactions. From the
experimental evidence it seems that the fluctuations
in the inverse compressibility
is proportional to $\langle \partial \mu/\partial N \rangle$
\cite{chang,marc,sba,cow}.
From numerical calculations one learns that for strong interactions 
the mean root square of the inverse compressibility are proportional to its
average\cite{sba}, with a proportionality constant of about $0.15$
for a very wide range of interactions and disorder strengths.
However, it is evident that for weak interactions
this ratio can not be constant.
Our goal is to study the behavior of the inverse compressibility
for weak interactions, and to understand the transition
to the strong interaction regime.

In this paper we extend the perturbative 
diagrammatic calculations to include e-e interactions.
The average compressibility, capacitance and inverse compressibility 
as a function of the interaction strength are calculated.
The fluctuations in the compressibility and capacitance
are found to be strongly suppressed
by the e-e interactions. The fluctuations in the the inverse compressibility
of the sample can 
also be deduced from the fluctuations in the compressibility. It turns
out that the fluctuations in the 
inverse compressibility, while not suppressed are relatively small.

This is checked by an exact diagonalization numerical
study of an interacting, spinless, disordered tight-binding model 
for nearest-neighbor interactions as well as for long-range Coulomb 
interactions. For weak interactions the numerical results
follow the perturbative calculation. As may be expected from earlier studies,
\cite{sba} deviations appear as the strength of interaction is increased.
Thus, a transition from weak-interaction behavior of
the fluctuations to strong-interaction behavior is evident.
In the end we discuss
the significance of these results for the interpretation of the experiments.

\section{Averaged Compressibility}

\begin{figure}
\centerline{\epsfxsize = 4in \epsffile{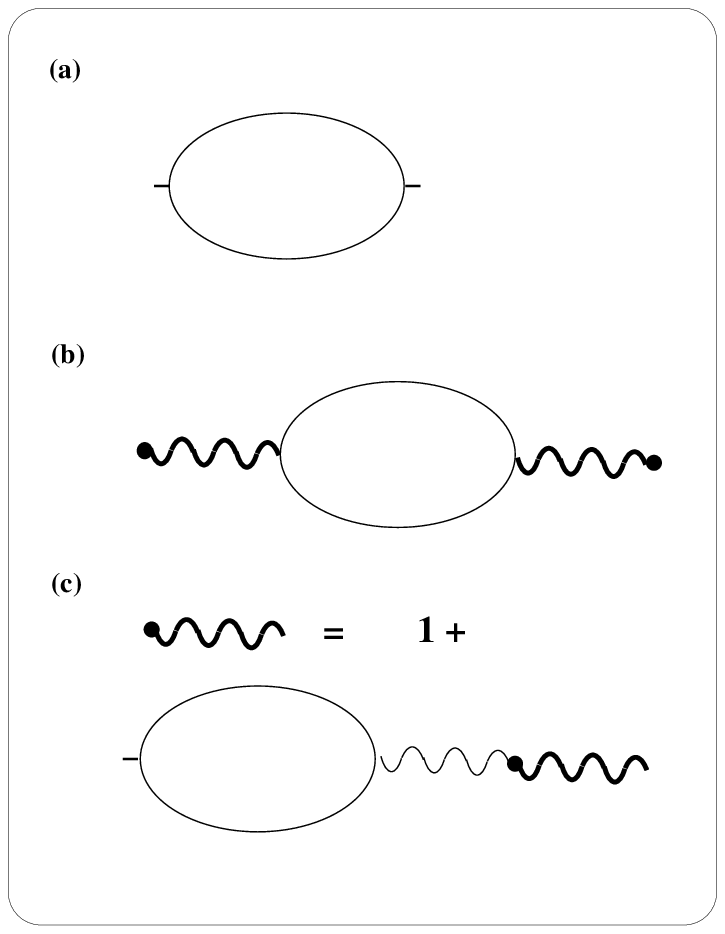}}
\caption {Feynman diagrams representing the average (a) non-interacting 
and (b) interacting compressibility., where the wavy lines correspond
to the electron-electron interaction given in diagram (c).
\label{fig.1}}
\end{figure}

The compressibility may be 
written in terms of the exact Green function of the system as:
\begin{equation}
{{\partial N} \over  {\partial \mu}} =
- {{1} \over {\pi}} \ {{\partial} \over  {\partial \mu}}
{\rm Im} \ {\rm Tr} \ \hat G^{R}(\varepsilon)
\label{e0}
\end{equation}
where $\mu$ is the Fermi energy of the system, $N$ is the number of
particles. Given the Hamiltonian $H$,
the exact Green function can be written as
\begin{equation}
\hat G^{R}(\varepsilon) = (\hat G^{A}(\varepsilon))^* = 
{{1}\over{\varepsilon - H - i 0}}.
\label{e0a}
\end{equation}
For the non-interacting case Eqs. (\ref{e0},\ref{e0a}) take the form
\begin{equation}
{{\partial N} \over  {\partial \mu}} =
- {{1} \over {\pi}} \ {{\partial} \over  {\partial \mu}}
{\rm Im} \int_{-\infty}^{\mu} d \varepsilon
\int_{\Omega} d \vec r G^{R}(\vec r,\vec r,\varepsilon),
\label{e1}
\end{equation}
where $\Omega$ is the volume and
\begin{equation}
G^{R}(\vec r, \vec r \ ',\varepsilon) =
\langle \vec r | \hat G^{R}(\varepsilon) | \vec r \ ' \rangle = 
\sum_n {{\psi_n(\vec r) \psi_n^{*}(\vec r \ ')} \over
{\varepsilon - \varepsilon_n - i 0}},
\label{e2}
\end{equation}
Here $\psi_n(\vec r)$ ($\varepsilon_n$) is the n-th eigenvector
(eigenvalue) of the system. This defines the 
density of states at the Fermi energy $\nu$ 
\begin{equation}
\bigg \langle {{\partial N} \over  {\partial \mu}} \bigg \rangle = 
\nu \Omega = {{1}\over{\Delta}}.
\label{e6}
\end{equation}
According to Eq. (\ref{e6}) in order to add one electron to
the system the Fermi energy should change by $\Delta$.
This can be also represented in a diagrammatic perturbation theory
by the diagram shown in Fig. \ref{fig.1}(a)
corresponding to
\begin{equation}
\bigg \langle {{\partial N} \over  {\partial \mu}} \bigg \rangle = 
{{i} \over {\pi}} \int_{-\infty}^{\mu} d \varepsilon
\int_{\Omega} d \vec r \langle G^{R}(\vec r, \vec r,\varepsilon)
G^{R}(\vec r, \vec r,\varepsilon)\rangle -
\langle G^{A}(\vec r, \vec r,\varepsilon)
G^{A}(\vec r, \vec r,\varepsilon)\rangle 
\label{e3}
\end{equation}
where $\langle \ldots \rangle$ denotes an average over different realizations
of disorder. 

Now let us discuss the influence of e-e interactions on the averaged
compressibility. One would expect that repulsive e-e interactions will reduce
the compressibility since it costs additional interaction energy to insert an
electron into the system. We begin by investigating the role
of short range interactions represented by 
\begin{equation}
U(\vec r, \vec r \ ')
= \lambda \nu^{-1} \delta (\vec r - \vec r \ ') =
a^d U \delta (\vec r - \vec r \ ')
\label{ez1}
\end{equation}
and 
\begin{equation}
U(\vec q) = \lambda \nu^{-1} = \lambda \Omega \Delta = a^d U,
\label{ez2}
\end{equation}
where $\lambda = e^2 \nu a^{d-1}$
is a dimensionless coupling constant and $U=e^2/a$;
here $a$ is the range of
the interaction and $d$ is the systems dimensionality. The compressibility
in the interacting case may be represented by the diagrams shown in Fig.
\ref{fig.1}(b), corresponding to
\begin{equation}
\bigg \langle {{\partial N} \over  {\partial \mu}} \bigg \rangle = 
\lim_{\vec q \rightarrow 0}
\nu \Omega \chi(\vec q) 
\label{e7}
\end{equation}
where $\chi(\vec q)$ is the result of summing the diagrammatic
series shown in Fig. \ref{fig.1}(c). For a short range interaction
\begin{equation}
\chi(\vec q) = {{1}\over{1+\lambda}}.
\label{e8}
\end{equation}
Inserting the result for $\chi(\vec q)$ into Eq. (\ref{e7}) one obtains
\begin{equation}
\bigg \langle {{\partial N} \over  {\partial \mu}} \bigg \rangle = 
\left({{1}\over{1+\lambda}}\right)
\left({{1}\over{\Delta}}\right) = 
\left({{1}\over{1+a^d U/ \Omega \Delta}}\right)
\left({{1}\over{\Delta}}\right).
\label{e9}
\end{equation}
Thus, the additional energy needed to add an electron due to the
e-e interactions (the charging energy) is $a^d U / \Omega$.
This is exactly the result expected
if one assumes a constant density of electrons in the system. 

The calculation of the compressibility for the long-ranged Coulomb
interaction can be performed along similar lines. For the Coulomb
interaction 
\begin{equation}
U(\vec r, \vec r \ ') = {{e^2} \over {|\vec r - \vec r \ '|}}
\label{ez3}
\end{equation}
and for an infinite system
\begin{equation}
U(\vec q) = \tilde S_d e^2 / q^{d-1}, 
\label{ez4}
\end{equation}
(where $\tilde S_2 = 2 \pi$ and
$\tilde S_3 = 4 \pi$) which for future convenience
in the numerical simulations may also be written as
$U(\vec r, \vec r \ ') = U a / |\vec r - \vec r \ '|$.
As can be seen in Eq. (\ref{e7}) the
zero mode ($\vec q = 0$) determines the behavior of the average
compressibility. Therefore, one must treat the zero mode carefully,
since simply inserting the infinite system value of the interaction
for $\vec q = 0$ will lead to no compressibility, i.e., infinite capacity.
The zero component of the Fourier transform for a finite
system is equal to
\begin{equation}
U_{\vec q=0} = {{1}\over{\Omega^2}} \int_{\Omega} d \vec r \ d \vec r \ '
{{e^2} \over {|\vec r - \vec r \ '|}} = {{S_d e^2}\over{L}}
\label{ez5}
\end{equation}
where
\begin{equation}
S_d = {{L}\over{\Omega^2}} \int_{\Omega} d \vec r \ d \vec r \ '
{{1} \over {|\vec r - \vec r \ '|}}
\label{ez6}
\end{equation}
is a numerical constant which depends on the geometry.
Thus repeating the summation
of the diagrammatic series shown in Fig. \ref{fig.1}(c)
one obtains
\begin{equation}
\chi(\vec q=0) = {{1} \over {1 + (\kappa L)^{d-1}}},
\label{e12}
\end{equation}
where $\kappa^{d-1} = S_d e^2 \nu$, which is
the usual random phase approximation (RPA) for the e-e interactions.
Using the above result in Eq. (\ref{e7}) results in
\begin{equation}
\bigg \langle {{\partial N} \over  {\partial \mu}} \bigg \rangle = 
\left( {{1}\over{\Delta}} \right) \left({{1}\over{1 + S_d a U / \Omega \Delta}}
\right) = \left( {{1}\over{\Delta}} \right)
\left({{1} \over {1 + (\kappa L)^{d-1}}}\right).
\label{e13}
\end{equation}
Therefore, the 
compressibility tends to its non-interacting value for $\kappa L \ll 1$
(i.e., large screening length, possible for example in 
very small semiconducting grains)
and to $(1/\kappa L)^{d-1}$ of its non-interacting value for $\kappa L \gg 1$
(metallic grains).

\section{Fluctuations in Compressibility}
\label{sf}

\begin{figure}
\centerline{\epsfxsize = 4in \epsffile{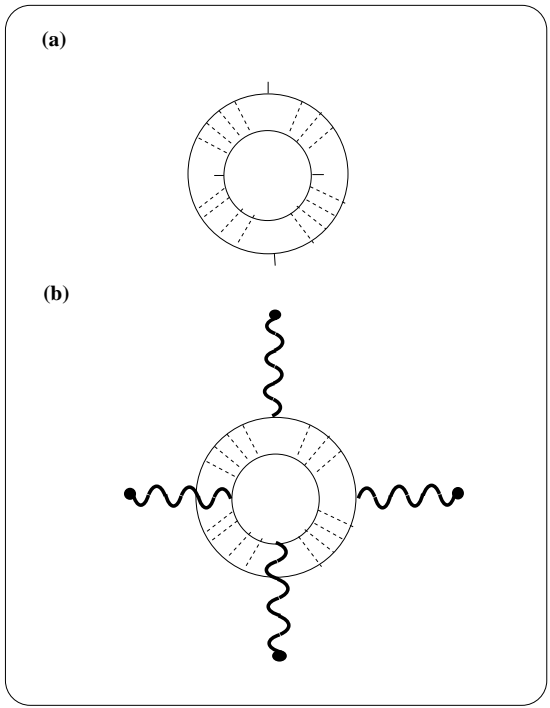}}
\caption {(a) Feynman diagrams representing the fluctuations in the 
compressibility of a sample. Diagram (a) corresponds to the non-interacting
case, while diagram (b) to the interacting case.
\label{fig.2}}
\end{figure}

In this section we shall calculate the fluctuations in the compressibility
in the presence of e-e interactions. For the non-interacting
electrons the fluctuations are expected follow the Wigner-Dyson statistics.
As was noted in Ref. \onlinecite{as}, in order to obtain the 
Wigner-Dyson statistics in the perturbative diagrammatic calculation
one must insert a cut-off in energy (or temperature) of order of the
single electron level spacing $\Delta$. Explicitly the fluctuations
can be calculated using the diagrams appearing in
Fig. \ref{fig.2}(a), which correspond to
\begin{equation}
\bigg \langle \delta^2 { {\partial N} \over  {\partial \mu}} \bigg \rangle = 
T \sum_m \sum_{\vec q} \omega_m {\cal D}_{\omega_m}^4 (\vec q).
\label{f1}
\end{equation}
where $\omega_m = 2 \pi m T$
is the Matsubara frequency and $T$ is the temperature.
${\cal D}_{\omega_m} (q)$
is the Fourier transform of the diffusion propagator which is the solution
to the equation
\begin{equation}
(\omega_m  - D \nabla^2) {\cal D}_{\omega_m}  (\vec r, \vec r \ ') =
D \delta  (\vec r - \vec r \ ') ,
\label{e4}
\end{equation}
with reflective boundary conditions $ \nabla_{\hat n}
{\cal D}_{\omega_m}  (\vec r, \vec r \ ') =0$, where $\hat n$ is the 
normal to the sample edge. For a rectangular grain of dimensions
$L^3$, the solution is
\begin{eqnarray}
{\cal D}_{\omega_m}  (\vec r, \vec r \ ') =
\frac {1}{\Omega} \sum_{n_i = -\infty}^\infty
\frac {\prod_{i=x,y,z} cos (k_i r_i) cos (k_i r_i ')}
{D q^2 + \omega_m},
\label{e5}
\end{eqnarray}
where $q^2 = k_{n_x}^2 + k_{n_y}^2 + k_{n_z}^2$, (for a 
two dimensional system the $\hat z$ component drops out) and
$k_i={\pi n_i}/{L}$.
After inserting the value of diffusion propagator given in Eq. (\ref{e5})
one is left with the following summation:
\begin{equation}
\bigg \langle \delta^2 { {\partial N} \over  {\partial \mu}} \bigg \rangle = 
T \sum_m \sum_{\vec q} \omega_m (\omega_m + D q^2)^{-4},
\label{f2}
\end{equation}
The most significant contribution comes
from the zero mode ($\vec q = 0$) resulting in
\begin{equation}
\bigg \langle \delta^2 { {\partial N} \over  {\partial \mu}} \bigg \rangle \sim
{{1}\over{\Delta^2}}.
\label{f3}
\end{equation}

The fluctuations in the interacting case are represented by the
diagrams shown in Fig. \ref{fig.2}(b) corresponding to
\begin{equation}
\bigg \langle \delta^2 { {\partial N} \over  {\partial \mu}} \bigg \rangle = 
T \chi^4(0) \sum_m \sum_{\vec q} \omega_m {\cal D}_{\omega_m}^4 (q),
\label{f5}
\end{equation}
where $\chi (0)$ is given by Eqs. (\ref{e8},\ref{e12}).
Note that $\chi$ appears in the power of four, which is the the result of the
fact that when two RPA lines intersect, four lines appear. 
For the short range interactions we obtain
\begin{equation}
\bigg \langle \delta^2 { {\partial N} \over  {\partial \mu}} \bigg \rangle \sim
\left({{1}\over{1+a^d U/ \Omega \Delta}}\right)^4 {{1}\over{\Delta^2}}.
\label{f6}
\end{equation}
Thus, the fluctuations in the compressibility are suppressed by the
interactions as compared to the non-interacting value. Even the relative
fluctuations (defined as the fluctuations in the compressibility divided 
by the averaged compressibility) are suppressed by the interactions.
According to Eq. (\ref{e9}):
\begin{equation}
\bigg \langle \delta^2 { {\partial N} \over  {\partial \mu}} \bigg \rangle 
\bigg/ \bigg \langle {{\partial N} \over  {\partial \mu}} \bigg \rangle ^2 \sim
\left({{1}\over{1+a^d U/ \Omega \Delta}}\right)^2.
\label{f7}
\end{equation}

A similar situation exists also for the Coulomb interactions.
By using the value of $\chi$
given in Eq. (\ref{e12}) one obtains
\begin{equation}
\bigg \langle \delta^2 { {\partial N} \over  {\partial \mu}} \bigg \rangle \sim
\left({{1} \over {1 + (\kappa L)^{d-1}}}\right)^4
{{1}\over{\Delta^2}},
\label{f8}
\end{equation}
and (using Eq. (\ref{e13}))
\begin{equation}
\bigg \langle \delta^2 { {\partial N} \over  {\partial \mu}} \bigg \rangle 
\bigg/ \bigg \langle {{\partial N} \over  {\partial \mu}} \bigg \rangle ^2 \sim
\left({{1} \over {1 + (\kappa L)^{d-1}}}\right)^2.
\label{f9}
\end{equation}
Thus, for $\kappa L \ll 1$ the relative fluctuations are the same as in the 
non-interacting case, while
$\kappa L \gg 1$ the relative fluctuations are suppressed by a 
factor of $(\kappa L)^{-2(d-1)}$.
It is important to note
that for most metallic and semiconducting samples the above relation
holds since $\kappa$ is of the order of the Fermi momentum, i.e.,
$\kappa L \sim 1$ for a grain with $\sim 1$ electron.

\section{Capacitance}

From the experimental point of view,
capacitance measurements is one of the
most accessible methods to investigate the spectral properties of quantum
dots and grains\cite{ash}. 
Therefore, it is interesting to connect the fluctuations
in the compressibility to fluctuations in the capacitance for interacting 
disordered systems. The
capacitance of a grain $C$ may be related to its chemical potential
in the following way:
\begin{equation}
{{e^2}\over{ C }} = 
{{\partial \mu} \over  {\partial N}} - \Delta,
\label{g1}
\end{equation}
which is equivalent to
\begin{equation}
C = {{1}\over{2^{d-1} \pi}} \Omega \kappa^{d-1} \left(
{{\partial \mu} \over  {\partial N}} - \Delta \right)^{-1},
\label{g2}
\end{equation}
when $\kappa L \gg 1$ and thus
one may write the average capacitance as
\begin{equation}
\langle C  \rangle = {{1}\over{2^{d-1} \pi}} \Omega \kappa^{d-1} 
\Delta \bigg\langle
{{\partial N} \over  {\partial \mu}} \bigg\rangle.
\label{g3}
\end{equation}
After inserting the calculated compressibility for the
long-range Coulomb interaction (Eq. (\ref{e13})) one obtains
\begin{equation}
\langle C  \rangle = {{1}\over{S_d}} L,
\label{g4}
\end{equation}
which is the expected purely geometrical value of the capacitance.

The fluctuations in the capacitance can be expressed via the fluctuations
in the compressibility in the following way
\begin{equation}
\langle \delta^2 C \rangle =
\left( {{1}\over{S_d}} \Omega \kappa^{d-1} \Delta \right)^2
\bigg\langle \delta^2 {{\partial N} \over  {\partial \mu}} \bigg\rangle,
\label{g5}
\end{equation}
which after inserting Eq.(\ref{f9}) results in
\begin{equation}
\langle \delta^2 C \rangle =
\langle C \rangle^2 \left( {{1}\over{\kappa L}}\right)^{2(d-1)}.
\label{g6}
\end{equation}
Thus when $\kappa L \gg 1$ the fluctuations in the capacitance are much 
smaller than the average capacitance.

\section{Inverse Compressibility}

As mentioned briefly in the introduction, the inverse compressibility
$\partial \mu/\partial N$ of a quantum dot can be deduced from
measurements of the spacings between consecutive gate voltages
for which the conductance through the dot peaks.
The spacings in the gate voltage are proportional to
the difference between the chemical potential of $N+1$ and $N$ electrons
\cite{sba} denoted by $\Delta^N_2$, which
may be related to the ground state energies in the following way:
\begin{equation}
\Delta^N_2= \mu_{N+1} - \mu_{N} = E_{N+1} - 2 E_N + E_{N-1},
\label{h0}
\end{equation}
where $E_N$ is the ground state
energy of the dot populated by $N$ electrons. In the continuum limit
$\Delta_2=\partial \mu/\partial N$, i.e., the experiment actually measures the 
discrete limit of the inverse compressibility. 

The first step in connecting the results obtained from the compressibility
calculations to the behavior of $\Delta_2$ is to relate 
$\langle \partial \mu/\partial N \rangle$ to
$\langle \partial N/\partial \mu \rangle$. 
With no loss of generality one can write
\begin{equation}
\bigg \langle {{\partial \mu} \over  {\partial N}} \bigg \rangle =
\left \langle \left( \bigg \langle {{\partial N} \over  {\partial \mu}} 
\bigg \rangle + \delta {{\partial N} \over  {\partial \mu}} \right)^{-1}
\right\rangle,
\label{h1}
\end{equation}
and 
\begin{equation}
\bigg \langle \delta^2 {{\partial \mu} \over  {\partial N}} \bigg \rangle =
\left \langle \left( \bigg \langle {{\partial N} \over  {\partial \mu}} 
\bigg \rangle + \delta {{\partial N} \over  {\partial \mu}} \right)^{-2}
\right\rangle - 
\bigg \langle {{\partial \mu} \over  {\partial N}} \bigg \rangle^2
\label{h2}
\end{equation}
which if one assumes a well behaved distribution of the compressibility
will result in
\begin{equation}
\bigg \langle {{\partial \mu} \over  {\partial N}} \bigg \rangle \sim
\bigg \langle {{\partial N} \over  {\partial \mu}}\bigg \rangle^{-1}.
\label{h3}
\end{equation}
In a similar manner
\begin{equation}
\bigg \langle \delta^2 { {\partial \mu} \over  {\partial N}} \bigg \rangle \sim
\bigg \langle \delta^2 { {\partial N} \over  {\partial \mu}} \bigg \rangle
\bigg \langle {{\partial N} \over  {\partial \mu}} \bigg \rangle^{-4}.
\label{h4}
\end{equation}
These relations are valid as long as 
$\langle \delta^2 \partial N / \partial \mu \rangle <
\langle \partial N / \partial \mu \rangle^2$ . From Eqs. (\ref{e9},\ref{e13})
and (\ref{f6},\ref{f8}) it can be seen that this condition is usually fulfilled
since $\kappa L \gg 1$. 
The above condition also holds in the non-interacting metallic regime for
which the usual diagrammatic expansion is valid.\cite{iga,krbg} 
Thus 
\begin{equation}
\bigg \langle {{\partial \mu} \over  {\partial N}} \bigg \rangle \sim
\left(1+{{a^d U}\over{ \Omega \Delta}}\right) \Delta
\label{h5}
\end{equation}
for short range interactions and
\begin{equation}
\bigg \langle {{\partial \mu} \over  {\partial N}} \bigg \rangle \sim
\left(1+\left(\kappa L\right)^{d-1} \right) \Delta =
\left(1+{{S_d a U}\over{L \Delta}}\right) \Delta
\label{h6}
\end{equation}
for Coulomb interactions. The fluctuations in both cases are equal to
\begin{equation}
\bigg \langle \delta^2 { {\partial \mu} \over  {\partial N}} \bigg \rangle \sim
\Delta^2.
\label{h7a}
\end{equation}
From Eqs. (\ref{h5},\ref{h6}) it is clear that $\langle
\partial \mu / \partial N \rangle$ grows proportionally to the interaction 
strength $U$,while the fluctuations are independent of the interaction 
strength.

This behavior implies that while the e-e interactions tend to shift 
the distribution of the inverse compressibility 
the width of the distribution will not change significantly\cite{fn1}.

The results for $\langle \partial \mu / \partial N \rangle$ could also
be anticipated from a simple assumption on the distribution of the
electron density of the ground-state. The
electrostatic energy needed to add an electron to a system of $N$ electrons 
is given by
\begin{equation}
\varepsilon_{int}^N = 
\int_{\Omega} d \vec r \ d \vec r \ ' \ U(\vec r , \vec r \ ') 
\rho_N(\vec r) \rho_1(\vec r \ '),
\label{h8}
\end{equation}
where $\rho_N$ is the density of the $N$ electrons already in the system,
and $\rho_1$ is the density of the additional electron.
Assuming that both densities are uncorrelated, and that
on the average $\langle \rho_N \rangle = N/\Omega$ and
$\langle \rho_1 \rangle = 1/\Omega$, the average
electrostatic energy needed to add an electron is
\begin{equation}
\langle \varepsilon_{int}^N \rangle = 
\int_{\Omega} d \vec r \ d \vec r \ ' \ U(\vec r , \vec r \ ') 
{{N}\over {\Omega^2}},
\label{h9}
\end{equation}
which for short-range interactions results in
$\langle \varepsilon_{int}^N \rangle = N a^d U / \Omega$ 
and for the Coulomb interactions
$\langle \varepsilon_{int}^N \rangle = 
N S_d e^2 L^{d-1} / \Omega = N (\kappa L)^{d-1} \Delta$.
Since under these assumptions
$\langle \partial \mu / \partial N \rangle = 
\langle \varepsilon_{int}^N \rangle - 
\langle \varepsilon_{int}^{N-1} \rangle + \Delta$,
one immediately obtains Eqs. (\ref{h5},\ref{h6}), which is the classical limit
of the Coulomb blockade.\cite{lk}

Also the fluctuations in the inverse compressibility may be deduced from
similar assumptions, resulting in
\begin{equation}
\langle (\varepsilon_{int}^N)^2 \rangle \sim
\langle \varepsilon_{int}^N \rangle^2,
\label{h11}
\end{equation}
for both short and long range interactions. Thus, the fluctuations in
the charging energy are
$\langle \delta^2 \varepsilon_{int}^N \rangle = 
\langle (\varepsilon_{int}^N)^2 \rangle -
\langle \varepsilon_{int}^N \rangle^2 \sim 0$,
This leads to the conclusion that there is no additional contribution 
to the inverse compressibility 
fluctuations beyond the fluctuations in $\Delta$, in agreement with Eq.
(\ref{h7a}).

Therefore, we expect that the results presented in the previous sections will
hold as long as the RPA assumptions hold, i.e., $v_f \gg  e^2/h$, where
$v_f$ is the Fermi velocity.

\section{Numerical Calculation of the Inverse Compressibility} 

In this section we shall numerically test the 
properties of the inverse compressibility. Of course, in a numerical
calculation we can only compute the discrete form of 
$\partial \mu / \partial N$, i.e, $\Delta_2$ as a function of the
strength of e-e interactions in the system.

As a model system we chose
a system of interacting electrons on a 2D cylinder
of circumference $L_x$ and height $L_y$, which has been previously used
in the study of the influence of e-e interactions on persistent currents.
\cite{pc} The model Hamiltonian is given by:
\begin{eqnarray}
H= \sum_{k,j} \epsilon_{k,j} a_{k,j}^{\dag} a_{k,j} - V \sum_{k,j}
(a_{k,j+1}^{\dag} a_{k,j} + a_{k+1,j}^{\dag} a_{k,j} + h.c)
+ H_{int}
\label{hamil}
\end{eqnarray}
where  $a_{k,j}^{\dag}$
is the fermionic creation operator,
$\epsilon_{k,j}$ is the energy of a site ($k,j$), which is chosen 
randomly between $-W/2$ and $W/2$ with uniform probability and $V$
is a constant hopping matrix element. $H_{int}$ is the interaction part of
the Hamiltonian given by:
\begin{equation}
H_{int} = {{U}\over{2}} \sum_{\{k,j;l,p\}} a_{k,j}^{\dag} a_{k,j}
a_{l,p}^{\dag} a_{l,p}
\label{hamil1}
\end{equation}
for the short range interactions (where $\{\ldots\}$ denotes nearest-neighbor
pair of sites) and
\begin{equation}
H_{int} = {{U}\over{2}}  
\sum_{k,j;l,p} {{a_{k,j}^{\dag} a_{k,j}
a_{l,p}^{\dag} a_{l,p}} \over 
{|\vec r_{k,j} - \vec r_{l,p}|/b}}
\label{hamil2}
\end{equation}
for the Coulomb interaction, where
$b$ is the lattice constant. 

For a sample of $M$ sites and $N$ electrons,
the number of eigenvectors spanning the many body Hilbert space
is $m = (_N^M)$. The many-body Hamiltonian may be represented by an
$m \times m$ matrix which is numerically diagonalized for different values
of the e-e interaction.
Here we consider a $4 \times 3$,
$4 \times 4$ and $4 \times 5$ lattices with $M=12,16,20$ sites 
correspondingly.
For each value of $M$ the average and fluctuations of $\Delta^{M/2}_2$
are calculated.
To obtain $\Delta^{M/2}_2$ in each case
$E_{gs}^N$ for $N=M/2-1$, $N=M/2$ and $N=M/2+1$ are calculated.
For the largest system considered ($M=20$,$N=10$)
the calculation of the ground-state energy
corresponds to diagonalizing a $184756 \times 184756$ matrix.
We chose $W=8V$ for which this system is in the metallic regime \cite{pc}
and average the results over $500$ realizations for each value
of interaction strength for the $M=12$ and $M=16$ cases and $200$ realizations
for $M=20$.

\begin{figure}
\centerline{\epsfxsize = 2.7in \epsffile{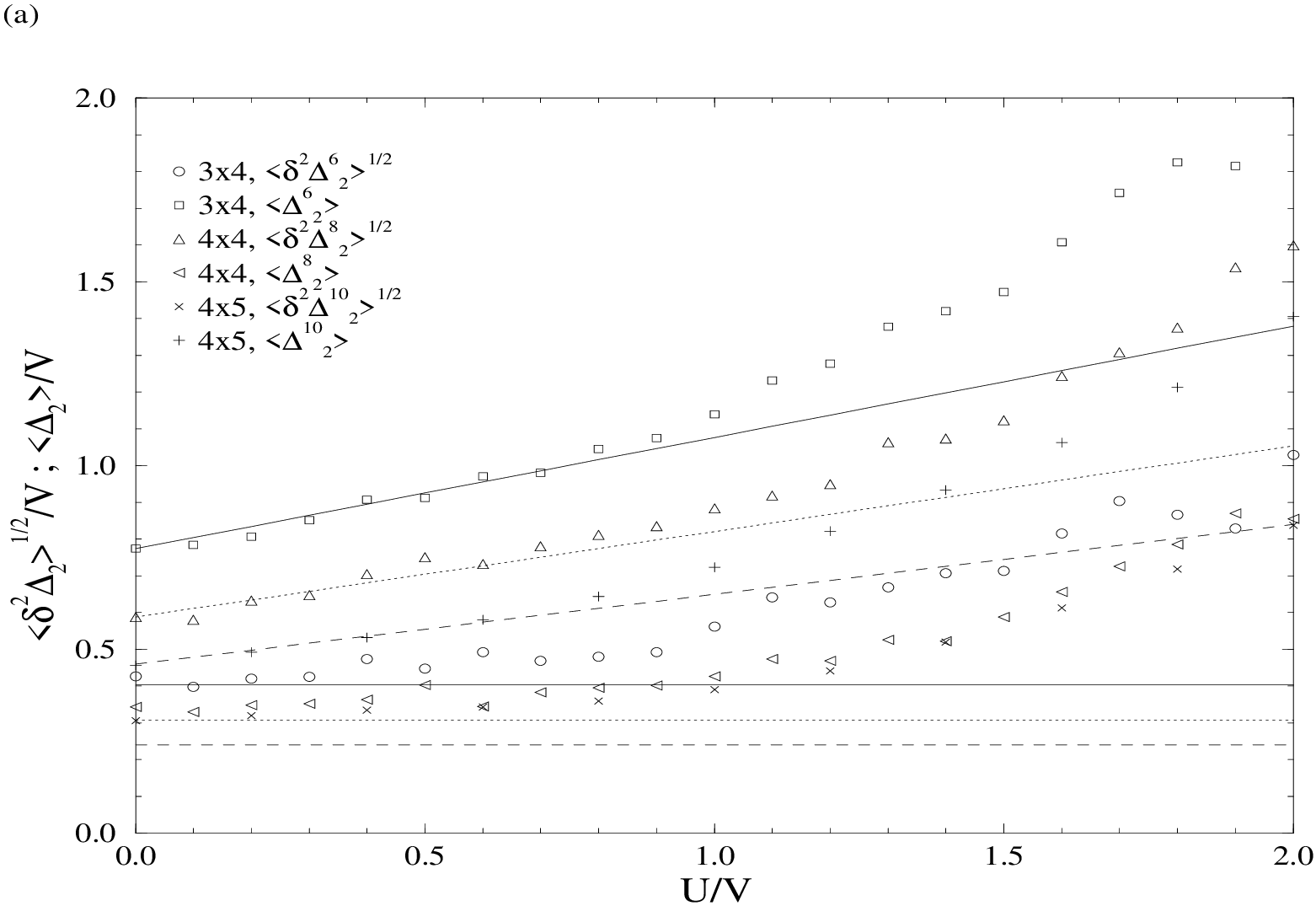}}
\centerline{\epsfxsize = 2.7in \epsffile{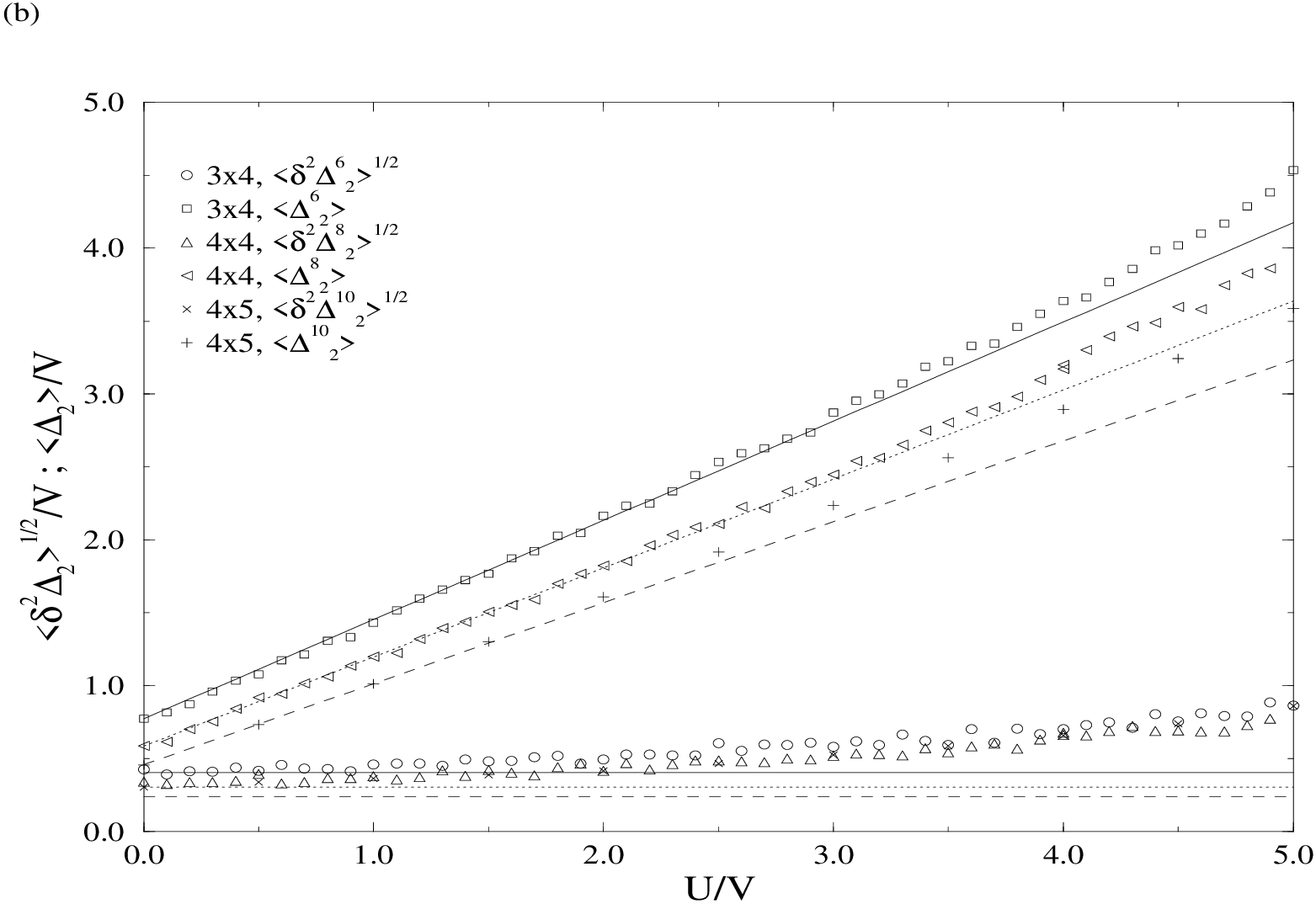}}
\caption {$\protect \langle \Delta^{M/2}_2 \rangle$ and 
$\protect \langle \delta^2 \Delta^{M/2}_2 \rangle$ for different lattice sizes
($3 \times 4$ , $4 \times 4$ and  $5 \times 4$) as function of interaction
strength for (a) short range interactions (b) Coulomb interactions.
The lines represent the theoretical predictions given in the text. The
full line corresponds to a $3 \times 4$ lattice, the dotted line to
$4 \times 4$ and the dashed line to $5 \times 4$.
\label{fig.3}}
\end{figure}

The results for $\langle \Delta^{M/2}_2 \rangle$ and 
$\langle (\delta^2 \Delta^{M/2}_2 \rangle)^{1/2}$ are given 
in Fig. \ref{fig.3}a
for the short-range interactions and in Fig. \ref{fig.3}b for the Coulomb
interaction. It can be seen that in both cases
$\langle \Delta^{M/2}_2 \rangle$ increases linearly for low values of $U$ while
$\langle \delta^2 \Delta^{M/2}_2 \rangle$ remains constant. 
This is in qualitative
agreement with Eqs. (\ref{h5} - \ref{h7a}). In order to obtain also
quantitative agreement one must take into account some finer details.
For $\langle \Delta^{M/2}_2 \rangle$ one must consider the fact that the 
calculations were performed on a lattice. For short range interactions,
one should replace $a^d / \Omega$ in Eq. (\ref{h5})
by $Z(M) / (M-1)$ where the
average number of nearest neighbors, $Z(M)$, depends on $M$ due to the
different ratio of sites close to the boundaries, where
for $M=12,16,20$, $Z(M)=3.333,3.5,3.6$ correspondingly.
Using the values of the single electron spacings $\Delta=0.42V,0.59V,0.77V$ for
$M=12,16,20$ in Eq. (\ref{h5}) results in the curves plotted
in Fig. \ref{fig.3}(a) for $\langle \Delta^{M/2}_2 \rangle$.
One can see a good fit up to $U=V$.
For the long range interactions in our lattice system
one should replace the integration in Eq. (\ref{ez6}) by a summation, i.e.
$S_2 = (L/M^2) \sum_{k,j \ne l,p} 
|\vec r_{k,j} - \vec r_{l,p}|^{-1}$, which
results in $S_2 = 2.35,2.44,2.48$
corresponding to $M=12,16,20$. After incorporating
$S_2$ and $\Delta$ in Eq. (\ref{h6}) one obtains
the curves plotted in Fig. \ref{fig.3}b.
An excellent fit is seen up to $U=3V$.
The exact value of $\langle \delta^2 \Delta^{M/2}_2 \rangle = 
(4/\pi - 1)\Delta^2$  is deduced from RMT and plotted in Fig. \ref{fig.3}.
It can be seen that this prediction is also confirmed for weak
interactions \cite{fn}.

\begin{figure}
\centerline{\epsfxsize = 4in \epsffile{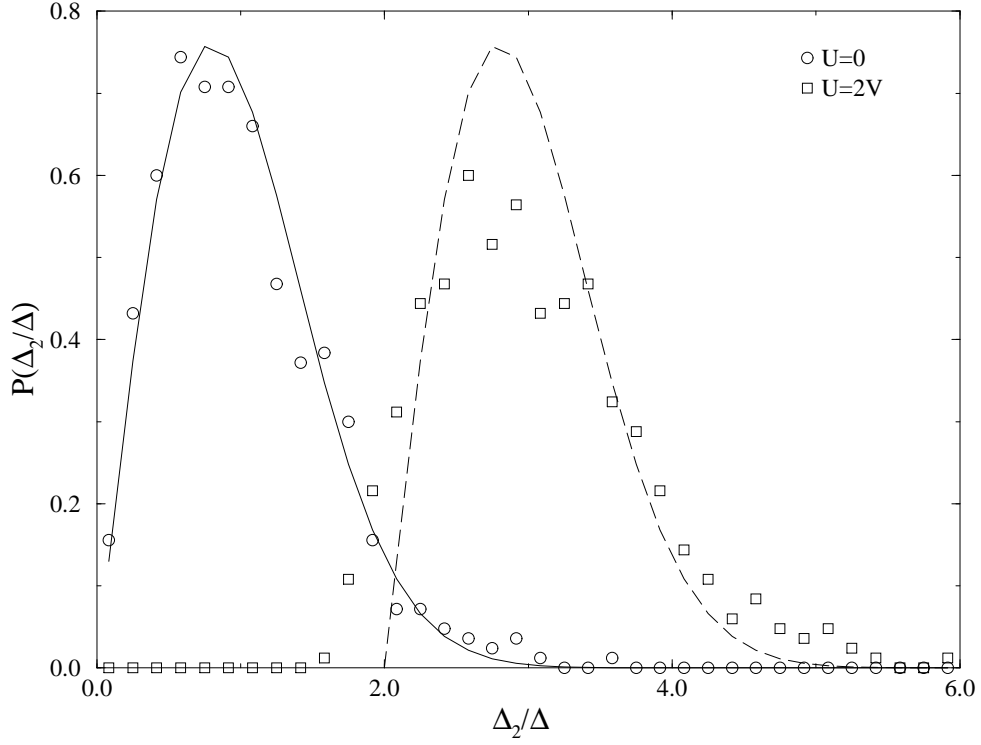}}
\caption {The full distribution $\protect 
P (\Delta^{M/2}_2/\Delta)$, 
for a $4 \times 4$ lattice.
The full line corresponds the GOE distribution
$\protect P(x)=(\pi x/2) \exp(-\pi x^2/4)$, where 
$\protect x = \Delta^{M/2}_2/\Delta$,
and the dashed line corresponds to the GOE distribution shifted
by the average charging energy for $U=2V$, i.e., by
$\protect \Delta^{M/2}_2 + \Delta - \langle \Delta^{M/2}_2 \rangle$.
\label{fig.4}}
\end{figure}

It is possible to observe what is the influence of the e-e interactions
on the full distribution of $\Delta_2$. In Fig. \ref{fig.4} the distribution
for the non-interacting case ($U=0$) is compared with the distribution for
$U=2V$ in the long range interaction case. As can be seen in 
Fig. \ref{fig.3}b, for this value of interaction strength no significant
change in the second moment of the distribution 
$\langle \delta^2 \Delta^{M/2}_2 \rangle$ is expected, and indeed the width of 
the distributions is almost identical. On the other hand, higher
moments seem to be influenced by the e-e interactions resulting in changes
in the tails of the distribution. The effect of interactions on the higher
moments of the $\Delta^{M/2}_2 $ merits further studies.

\begin{figure}
\centerline{\epsfxsize = 4in \epsffile{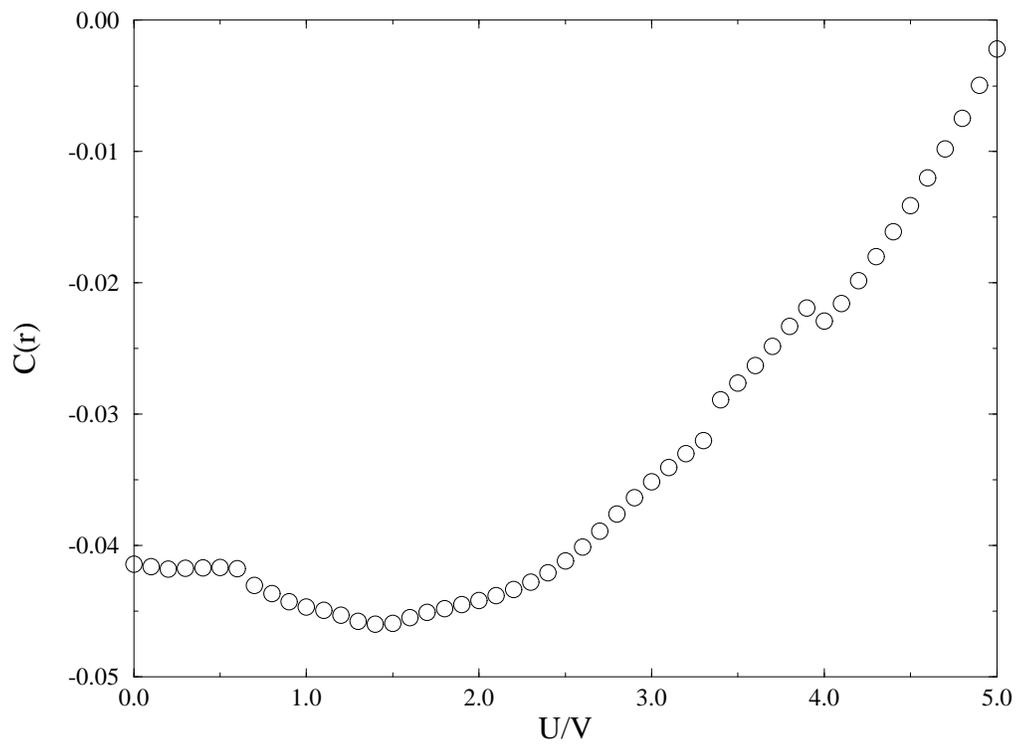}}
\caption {The two point correlation $\protect C(r= 2^{1/2} b)$,
for long range interactions, as a function of the interaction strength. 
\label{fig.5}}
\end{figure}

As has been mentioned in Ref. \onlinecite{sba}, and discussed in the previous 
section, we expect the analytical results based on the RPA approximation
to hold as long as no correlations develop in the electron density. 
Following Ref. \onlinecite{sba} we define a two point correlation function:
\begin{equation}
C(r) =  {{\sum_{k,j>l,p} C(\vec r_{k,j} - \vec r_{l,p} ) 
\delta_{|\vec r_{k,j} - \vec r_{l,p}|, r}} \over 
{\sum_{k,j>l,p} \delta_{|\vec r_{k,j} - \vec r_{l,p}|, r}}}\ ,
\label{corr}
\end{equation} 
where 
\begin{equation}
C(\vec r_{k,j} - \vec r_{l,p} ) = \left\langle
{{[a_{k,j}^{\dag} a_{k,j} - \langle a_{k,j}^{\dag} a_{k,j} \rangle]
[a_{l,p}^{\dag} a_{l,p}  - \langle a_{l,p}^{\dag} a_{l,p}  \rangle ]} 
\over 
{\langle a_{k,j}^{\dag} a_{k,j} \rangle
\langle a_{l,p}^{\dag} a_{l,p}  \rangle}}
\right\rangle.
\label{corr1}
\end{equation}
In Fig. \ref{fig.5} we present the correlation $C(r=\sqrt{2}b)$ (which
corresponds to a pair of diagonal sites)
for the long-range interactions as function of the interaction strength. 
Under the RPA assumptions
we expect $C(r=\sqrt{2}b) \rightarrow 0$. As can be seen 
there is some correlation as result of the boundaries even without e-e 
interactions. As the interaction strength increases there is no dramatic
change in $C(r=\sqrt{2}b)$
up to $U=2.5V$ and then one sees a strong deviation.
A similar behavior can be seen for the short range interactions
where the deviation appear at $U=V$. This agrees well with the values
of interaction for which the numerical results depart from the
RPA analytical calculations.

One can roughly estimate the strength of interaction $U$ for which
these correlations should appear. 
As mentioned in the previous section, we expect the RPA approximation
to hold while $e^2/v_f < 1$, which for long range interactions corresponds to
$r_s=\sqrt{\pi/2} (U / 4V) < 1$ and for short range interactions to
$r_s=\sqrt{\pi/2} (Z U / 4V) < 1$. This is surprisingly close to the
values for which deviations from RPA theory are seen numerically.

\section{Discussion}

The main conclusion that follows 
of the previous sections is that the RPA approximation
fails to explain the large fluctuations in the inverse compressibility
seen in the experiment \cite{chang,marc,sba}.
Under the assumptions of an RPA treatment of the e-e interactions
the fluctuations are proportional to the single electron mean level spacing,
therefore independent of the strength of the e-e interactions.
In contrast, experiments show that the fluctuations are substantially 
larger than the mean level spacing, and seem to be about
15 percent of
the average inverse compressibility.

Actually, one could have anticipated the failure of the RPA calculation
for the experimental realizations, since their densities are too low for
the RPA approximation to remain valid. For a 2DEG quantum dot 
one may rewrite the condition for which
the RPA fails $r_s=e^2/v_f<1$ as $n<1/\pi a_B^2$, where $n$ is the
electron density in the dot and $a_B$ is the Bohr radius, which in GaAs
is $\sim 100 \AA$. Thus one expects RPA to fail at densities
$n<3 \times 10^{-11} {\rm cm}^{-2}$. In all of the recent experiments for which
large fluctuations were observed \cite{chang,marc,sba} the 2DEG
density is about $3 \times 10^{-11} {\rm cm}^{-2}$, and the density in the dot
is probably even lower. 

The large fluctuations in the inverse compressibility are caused by spatial
fluctuations in the electron density which are not taken into account 
in the RPA approximation \cite{sba}. It is important to note that although 
the charge distribution becomes inhomogeneous when
the electron density is still much higher than the
one critical for Wigner crystallization. 
Thus the experiments are apparently in an intermediate range of densities 
(or intermediate e-e interaction strength). In this regime 
RPA approximation does not hold anymore and
some density correlations do appear, however the systems are still very far 
from the Wigner crystallization.

Note that in the case of large dimensionless conductance $g \gg 1$
that was considered in this paper we found no influence of the disorder on
the structure of Coulomb blockade peaks. For different physical reasons
fluctuations of the order of the 
inverse compressibility may also appear
in the localized regime.\cite{shk} 

Finally, it is also interesting to compare the fluctuations in 
the compressibility
to the fluctuations in the polarizability.\cite{ba} Although the calculation
of both quantities show many similar features, there is one crucial difference.
The main contribution to the fluctuations in the compressibility comes
from the zero-mode, which corresponds in the non-interacting case to
the single electron level fluctuations. In the polarizability, there
is no contribution from the zero mode, and the fluctuations stem from the
other modes, which correspond to density fluctuations. This
results in the fact that the relative
polarization fluctuations are smaller than the relative compressibility
fluctuations by a factor $1/g^2$.

\section*{Acknowledgments}

We would like to thank A. Auerbach, M. Field, Z. Ovadyahu,
B. Shklovskii, U. Sivan and D. A. Wharam for useful discussions.
R.B. would like to thank the US-Israel Binational Science Foundation
for financial support.

\begin{references}
\bibitem{fn1}
On the other hand, since we did not calculate higher moments, it is
possible that the tail of the distribution will change as function of
the interaction strength.
\bibitem{fn} Some deviations occur for $M=20$, probably because for this
larger lattice size the system is closer to the Anderson transition.
\end{references}
\end{document}